# CNNs and GANs in MRI-based cross-modality medical image estimation

Azin Shokraei Fard[1], David C. Reutens[1,2] and Viktor Vegh[1,2,*]

[1]Centre for Advanced Imaging, University of Queensland, Brisbane, Australia

[2]ARC Centre for Innovation in Biomedical Imaging Technology, Brisbane, Australia

[*]Corresponding author, e-mail: v.vegh@uq.edu.au

# Abbreviations used

$^{18}$F-FDG Fluorodeoxyglucose (18F)

ALMedian Median Value of Atlas Images

ALWV Atlas-based Local Weighted Voting

ALWV-Iter Iterative Atlas-based Local Weighted Voting

CC Correlation Coefficient

Convolutional neural networks (CNNs)

CAE-GAN Conditional Auto-Encoder GAN

cGAN Conditional GAN

DSC Dice Similarity Coefficient

DCNN Deep CNN

DECNN Deep Embedding Convolutional Neural Network

DSC Dice Similarity Coefficient

DECT Dual Energy CT

FCN Fully Connected Network

FCM Fuzzy C-Means

GPU Graphics Processing Unit

Generative adversarial networks (GANs)

HU Hounsfield Unit

LSGAN Least Square GAN

LA-GAN Locality-adaptive GAN

MAE Mean Absolute Error

MSE Mean Squared Error

MPRAGE Magnetization Prepared Rapid Acquisition Gradient Echo

MLEM Maximum Likelihood Expectation Maximisation

NMSE Normalised Mean Squared Error

NRMSE Normalised Root Mean Squared Error

ND: Not Disclosed

PSNR Peak Signal-to-Noise Ratio

PCC Pearson Correlation Coefficient

ReLU Rectified Linear Unit

ResNet Residual Network

SNR signal-to-noise ratio

SSIM Structural Similarity Index Measure

UTE Ultrashort Echo-Time

UNIT Unsupervised Image-to-image Translation

VAE Variational Autoencoder

ZeDD-CT  Zero-Echo-Time and Dixon Deep Pseudo-CT

## Abstract

Cross-modality image estimation involves the generation of images of one medical imaging modality from that of another modality. Convolutional neural networks (CNNs) have been shown to be useful in identifying, characterising and extracting image patterns. Generative adversarial networks (GANs) use CNNs as generators and estimated images are discriminated as true or false based on an additional network. CNNs and GANs within the image estimation framework may be considered more generally as deep learning approaches, since imaging data tends to be large, leading to a larger number of network weights. Almost all research in the CNN/GAN image estimation literature has involved the use of MRI data with the other modality primarily being PET or CT. This review provides an overview of the use of CNNs and GANs for MRI-based cross-modality medical image estimation. We outline the neural networks implemented, and detail network constructs employed for CNN and GAN image-to-image estimators. Motivations behind cross-modality image estimation are provided as well. GANs appear to provide better utility in cross-modality image estimation in comparison with CNNs, a finding drawn based on our analysis involving metrics comparing estimated and actual images. Our final remarks highlight key challenges faced by the cross-modality medical image estimation field, and suggestions for future research are outlined.

# 1. INTRODUCTION

Medical imaging is the process of generating images of body parts and organs to help diagnose diseases and disorders, plan treatments and monitor disease progression. The different types of medical imaging devices deliver complementary information. Magnetic resonance imaging (MRI) provides information on soft tissues, computed tomography (CT) is mostly used to image high electron density tissues such as bone but can also provide a level of soft tissue contrast, and molecular imaging techniques, namely single photon emission computed tomography (SPECT) and positron emission tomography (PET), image specific biological functions using radiotracers. Multi-modality medical images are now integral to clinical decision making. However, acquisition of images using multiple modalities has distinct disadvantages because multiple scans increase cost and radiation dose and delay clinical workflow. Cross-modality image estimation has the potential to overcome specific, existing limitations.

Cross-modality image estimation involves the synthesis of one image modality images from images acquired using a different modality. Benefits of cross-modality image estimation from a patient perspective include fewer scans, less delay, and lower radiation dose and healthcare costs. The medical provider benefits also by having faster scanning turnaround and higher patient throughput, reduced staffing and reduced healthcare and maintenance costs, on top of lower radiotracer production runs. Registration mismatch across modalities is also eliminated with cross-modality image estimation, as images are estimated within the same coordinate and orientation system. Whilst these are important advantages of cross-modality image estimation, careful consideration should be made towards the validity of image estimation for clinical use. This is because the physics governing image formation for the different modalities is fundamentally distinct. MRI has been a primary modality in cross-modality image estimation, as it is non-invasive and does not present radiation exposure to patients.

Multi-modality medical images came to feature in routine clinical practice around the early 1990s, and cross-modality image estimation methods have been of interest since about 2010. Image estimation methods fall into three categories – segmentation, atlas and machine learning based. Machine learning based image estimation has only been popular recently, as the tools continue to be developed. In the medical imaging field, supervised machine learning methods, particularly regression techniques, have been employed for image estimation. The convolutional neural network (CNN) within the deep learning framework has increasingly been used for multi-modality image estimation with mounting success. This review provides an overview of recent achievements in CNN-based image estimation, how the field has progressed towards generative adversarial networks (GANs) based on CNN constructs, and how these methods benchmark against other image estimation techniques. Appendix A provides the process used for literature selection and summary of papers reviewed. Appendices B and C respectively outline the technical aspects of the CNN and GAN frameworks for image estimation.

## 1.1 THE NEED FOR CROSS-MODALITY IMAGE ESTIMATION

According to methods reported in the literature, the application of CNNs/GANs for cross-modality image estimation fall into three primary groups. These include the synthesis of the CT and PET image from MRI, and the estimation of high resolution and high signal-to-noise ratio (SNR) PET images from low resolution and low SNR PET images using MRI as an anatomical high resolution reference. Other combinations have also been considered, but to a much smaller extent and generally without a clear justification.

### 1.1.1 CT image synthesis from MRI

The superior soft tissue contrast and avoidance of radiation exposure has led to MRI replacing CT in many clinical applications. However, CT images still have to be acquired for radiotherapy treatment dose calculation and for PET attenuation correction to be able to generate anatomically accurate images (Leynes et al. 2018). The increased uptake of PET-MRI hybrid systems led to other investigations in which MRI was used to estimate attenuation correction maps normally derived from CT scans. Essentially, the relationship between the CT image and attenuation map is a transformation from Hounsfield units to a linear attenuation coefficient. The attenuation coefficient in high electron density tissues such as bone is large, while in soft tissue it is a small number. In CT bone appears hyperintense and in MRI hypointense or as a null signal, but soft tissue contrast around bone in MRI provides a mechanism by which bone patterns can be delineated.

### 1.1.2 Estimating the PET image from MRI

It can be difficult to acquire imaging data in certain patient cohorts. Instances arise in the clinical setting where patients who have undergone MRI scans were not able to complete a PET scan. An example of such a case is Alzheimer's disease (Li et al. 2014a). Here, MRI can be considered for estimating $^{18}$F-FDG-PET images, as it is known that the $^{18}$F-FDG tracer is taken up by the brain in general, resulting in PET soft tissue anatomical contrast. This formed the basis for MRI, a soft tissue contrast imaging modality, to be used for PET image estimation. It is important to note that to perform cross-modality image estimation, a level of information correspondence has to exist between the imaging modalities. Hence, PET images of highly specific radiotracers with little or no anatomical image contrast are unlikely predictable from soft tissue contrast MRI scans.

### 1.1.3 Resolution and SNR boost in PET using MRI

PET images are inherently lower resolution than MRI scans due to the nature of the medical imaging modality. The 2-3mm to 1mm mismatch in image resolution provides several disadvantages. These include anatomical and functional mismatch and blurring between the modalities, and reduced specificity of regions identified using PET, especially important in oncology and neurology (Song et al. 2020a). The approach to bridge this gap has been to incorporate anatomical priors from MRI within a machine learning framework with the outcome of estimating PET images which match the MRI resolution.

Dosimetry in PET is an important factor which influences image quality and SNR. It is desirable to reduce the PET dose given to a patient due to risks associated with radiation exposure (Xiang et al. 2017). However, low dose PET studies inherently lead to low SNR PET images. Whilst image SNR can be increased by raising the radiotracer dose, the preferred methods in low-dose PET studies are post-processing approaches for noise reduction. As such, machine learning approaches which take noisy images as input and produce low noise images as an output have been investigated.

### 1.2 Metrics to assess whether the need can be met

Three metrics are generally used in evaluating the quality of image estimation. These include mean absolute error (MAE), mean squared error (MSE) and peak signal-to-noise ratio (PSNR):

$$\text{MAE} = \frac{1}{N} \sum_{n=1}^{N} |I(n) - I_e(n)|,$$

$$\text{MSE} = \frac{1}{N}\sum_{n=1}^{N}(I(n) - I_e(n))^2,$$

$$\text{PSNR} = 10\log_{10}\frac{MAX^2}{MSE},$$

Where $N$ is the total number of pixels in an image, $I(n)$ is an image intensity in the original image at location $n$ and $I_e(n)$ is the corresponding estimated value, and MAX is the maximum pixel value in the image data. The dice similarity coefficient (DSC) and Pearson correlation coefficient (PCC) have also been used, which measure how well the estimated image pixel intensities correlate with those in the expected image. The smaller the MAE and MSE and larger the DSC and PCC the better the quality of the image estimation. These quality metrics will be referred to in this review.

The reader is referred to Appendix B for specific terms in relation to the CNN architecture, and their extension to the GAN framework is provided in Appendix C.

## 2. MRI-BASED ESTIMATION OF CT

Table 3 and Table 4 summarise the literature and key parameters along with error measures for CNNs and GANs, respectively.

### 2.1 CNN implementations

Nie et al. proposed a 3D fully connected CNN with three convolutional layers (Nie et al. 2016a) based on the Caffe architecture (Jia et al. 2014). They did not implement pooling to be able to preserve the spatial resolution of the feature maps in the network used for CT image estimation. Three activation functions were evaluated - Relu, Sigmoid and Tanh. The network was designed to take image patches as input. MRI and CT pelvic datasets were acquired from 22 subjects from which patches mostly covering the 3D image volume were extracted for training the network. They found the 3D CNN was able to overcome apparent discontinuities across slices associated with 2D CNN implementations. They also found the PSNR with the Relu activation function to be larger than when Sigmoid or Tanh activation functions were used.

Roy et al. (Roy, Butman, and Pham 2017) generated synthetic CT images using two different echo time ultrashort echo-time (UTE) MRI data, since it has been established that a reduction in echo time can lead to sensitivity to a bone signal. The network adopted inception blocks via GoogleNet (Szegedy et al. 2015). This allowed 3D patches from two different echo time UTE inputs to be first processed through individual inception blocks before concatenation, followed by another inception block. Their result showed that the synthetic CT produced by the network with two UTE inputs led to larger PSNR in the estimated CT image than when MPRAGE, the traditional T1-weighted MRI data, was used as the input to the CNN. Notably, the second UTE sequence echo time was not actually in the ultra-short regime and it was close to that in the MPRAGE sequence (2.46ms versus 3.03ms), suggesting that the inclusion of ultra-short echo time data (70µs) via channel 1 provided the PSNR improvement. The question of using only the shortest echo time data was not considered.

Xiang et al. (Xiang et al. 2018) proposed a deep embedding convolutional neural network (DECNN) for synthesising the CT image from MRI data. Their intention for the 3D patch-based implementation was to overcome single slice processing used in 2D versions, the adverse outcome of which has been signal artifacts across slices in estimated images when stacked into 3D volumes. The network involved five consecutive convolutional layers through which MRI data was processed, before going through embedding blocks. Each embedding block consisted of two convolutional layers followed by one concatenation layer, generating a total of 131 feature maps. After each embedding block another convolutional layer was implemented for feature map refinement. Xiang et al. synthesised the CT image from MRI in both brain and prostate regions based on a sample size of 16 and 22, respectively. According to brain and prostate synthetic CT image findings, DECNN performed best in terms of mean PSNR and mean MAE. The comparison was made with respect to Atlas-based, random forest (a non-CNN learning method), CNN and fully connected network (FCN) implementations, see Table 1.

Table 1. Provided are the mean MAE and mean PSNR of the CT synthetic results using state-of-the-art methods. Reproduced from work by (Xiang et al. 2018).

| Remark | Atlas | RF | CNN | FCN | DECNN | Atlas | RF | CNN | FCN | DECNN |
|---|---|---|---|---|---|---|---|---|---|---|
| | Brain | Brain | Brain | Brain | Brain | Prostate | Prostate | Prostate | Prostate | Prostate |
| Mean MAE | 169.5 ± 35.7 | 99.9 ± 14.2 | 93.6 ± 11.2 | 88.9 ± 10.6 | 85.4 ± 9.24 | 64.6 ± 6.6 | 48.1 ± 4.6 | 45.3 ± 4.2 | 42.4 ± 5.1 | 42.5 ± 3.1 |
| Mean PSNR | 20.9 ± 1.6 | 26.3 ± 1.4 | 26.7 ± 1.0 | 27.1 ± 1.3 | 27.3 ± 1.1 | 29.1 ± 2.0 | 32.1 ± 0.9 | 33.1 ± 0.9 | 33.4 ± 1.1 | 33.5 ± 0.8 |

### 2.1.1 Application of the U-Net CNN

Han (Han 2017) used brain T1-weighted MRI and corresponding CT images from 18 patients for CT image estimation using a 2D deep CNN (DCNN), where 2D refers to an image slice. They modified a previously described U-Net architecture (Ronneberger, Fischer, and Brox 2015) by removing the 3-fully connected layers, which allowed 90% reduction in the number of parameters and resulted in faster CNN training. Han found the slice-by-slice estimation to be more efficient than using 3D image volumes, since more training data is available when 3D image volumes are converted to 2D slices and the computational overhead and GPU memory requirements are reduced as well. The overall MAE and MSE were smaller using the DCNN when compared with the Atlas-based method, and the PCC was larger as well, refer to Table 2.

Table 2. Han.'s MAE, MSE and PCC metrics for CT estimation form MRI using the DCNN architecture compared with the Atlas-based result.

| Quality metric | DCNN | Atlas-based |
|---|---|---|
| MAE | 84.8 | 94.5 |
| MSE | 188.6 | 198.3 |
| PCC | 0.906 | 0.896 |

Arabi et al. (Arabi et al. 2018) implemented Han's model and tested it using a cohort of 39 patient pelvic region datasets for estimating the CT image from T2-weighted MRI. They compared the estimated results with atlas-based methods, including ALMedian, (Sjölund et al. 2015) ALWV, (Dowling et al. 2015) ALWV Bone, (Arabi et al. 2016) and ALWV-Iter (Burgos et al. 2017). The CNN image estimation approach could, in most cases, achieve the best MAE across the different tests. Particularly, DSC using the CNN was the largest in the bladder, rectum and bone, but for the body and bone-thresh (i.e. segmented bone with the intensity threshold of 140 HU) two atlas-based methods performed better. The training time for the 2D model was about 2.5 days and CT estimation of axial slices took approximately 9s (Arabi et al. 2018).

Chen et al. (Chen, Qin, et al. 2018) also adapted a standard U-Net architecture for estimating the CT image from T2-weighted MRI of the prostate based on 51 patient datasets. To improve training, the U-Net was modified by implementing batch normalisation immediately after each convolutional layer and prior to the activation function. Moreover, they used up-sampling via zero padding for each convolutional layer to preserve the size of the feature maps after convolution. Comparison between the proposed U-Net and an atlas-based method (Chen et al. 2016) found that the image estimation time using U-Net was 3.84 to 7.65s, and on the same GPU 2 to 3 minutes for the atlas-based method. The body and soft tissue MAE were smaller using the CNN than that produced by the atlas-based method, and in bony structures the error was larger.

In the same year Fu et al. (Fu et al. 2019) adopted Han's model (Han 2017), and used 20 pelvic scans from patients with prostate cancer. They applied distinct modifications to the CNN by replacing the unpooling layers with fractionally-strided convolutional layers (similar to transpose convolution) to produce dense, high-resolution feature maps, instead of sparse, high-resolution ones normally achieved via unpooling layers. They additionally implemented the residual shortcuts proposed for the ResNet architecture (He et al. 2016), leading to encoder feature maps (before downsampling) which correspond with upsampled feature maps (decoder) in the residual frame. This change allowed for more rapid convergence during training. To compensate for the spatial resolution reduction in the deconvolutional layer, instance normalisation layers were applied instead of batch normalisation to upsample feature maps. Here, the expansion from a 2D to a 3D model was described as well, essentially replacing 2D operations with 3D equivalents. They used the Adam stochastic gradient descent method to minimise the loss function. At each iteration, a mini batch of 2D or 3D training images were randomly selected from the training set. Since the 3D model requires larger memory resources, a mini batch of 15 training slices and 1 training volume were introduced to run 2D and 3D CNN models. The results show that the training times for the 2D and 3D networks were approximately 4 and 2 hours, respectively, and 2D and 3D image estimation times were consistently around 5.5s. The average MAE across all tests were smaller in the case of 3D images (37.6) compared with 2D images (40.5). Example synthesised images are provided in Figure 1, and image metrics in Figure 2.

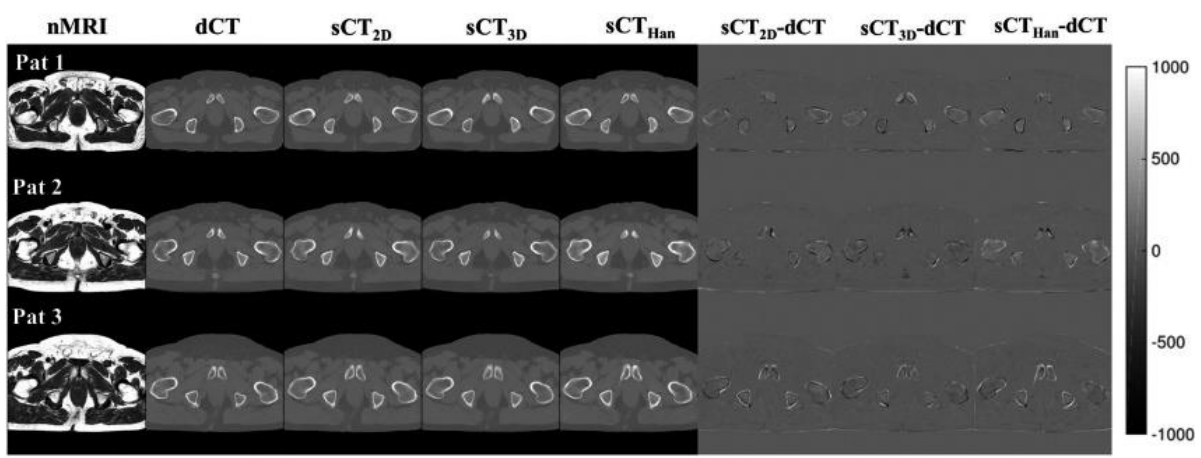

*Figure 1 Shown are real T1-weighted MR and real and estimated CT images from 3 patients. From left, normalized MRI images (input); corresponding deformed CT images (ground-truth); estimated CT images (using 2D network); estimated CT images (using 3D network); and estimated CT images (using Han's model). The difference maps between real and estimated CT using each network (i.e., proposed 2D, 3D and Han's) is shown in the last three columns. Adapted from (Fu et al. 2019).*

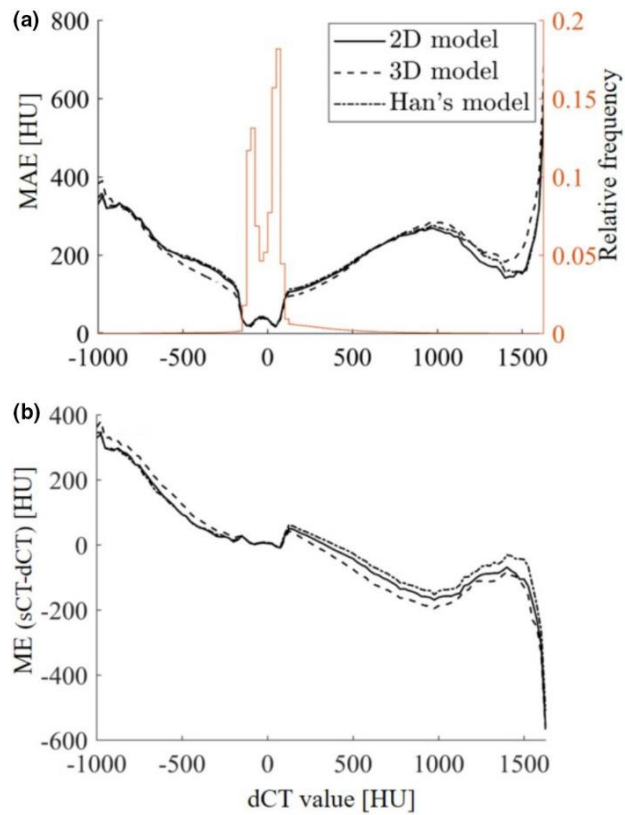

*Figure 2 MAE (a) and (ME) of voxels calculated within body masks for estimating sCT from MRI using three models: Fu's 2D model, Fu's 3D model, and Han's 2D model. Reproduced from (Fu et al. 2019).*

Leynes et al. (Leynes et al. 2018) used a 3D U-Net for estimating CT images in the pelvic region based on MRI data, i.e. zero echo time (ZTE) and Dixon water-fat scans, and they called the method ZTE and Dixon deep pseudo-CT (ZeDD-CT). The network required three inputs, namely ZTE images along with Dixon fat and Dixon water images. The 13-layer CNN was trained using 10 patient datasets and 16 datasets were used for evaluation. In particular, the RMSE in 30 bone lesions and 60 soft tissue lesions were assessed. They showed ZeDD-CT to outperform Dixon-CT for PET attenuation correction, suggesting the additional use of zero echo time MRI data provided better bone estimation, a key requirement of attenuation correction maps.

For radiotherapy treatment planning in oncology, Gupta et al. (Gupta et al. 2019) developed a 2D U-Net to estimate the CT image from MRI using sagittal head images (47 for training and 13 for testing). The input to the CNN were 2D T1-weighted MRI slices producing three output channels in Hounsfield units, corresponding with air, soft tissue and bone. This allowed the authors to differentiate the tissue intensity in the three primary tissue classes, as necessary for radiotherapy treatment planning. Spadea et al. (Spadea et al. 2019) also proposed a U-Net inspired 2D DCNN based on Han's model (Han 2017). In this study 3D T1-weighted MRI and 3D CT images were sliced into three groups, corresponding with axial, sagittal and coronal orientations. A DCNN was trained for each orientation. The median of the three outputs was used to generate the synthetic CT image in Hounsfield units. The DCNNs were trained using 12 patient datasets, tested and validated using an additional 1 and 2 datasets, respectively.

Training of the network took approximately 60 hours and CT image estimation took less than 30 seconds.

The previous best DCNN result in the field was reported based on dilated CNN using 2-fold cross validation (Dinkla et al. 2018). The dilation of convolutional filters was used to replace the down-sampling layers (i.e., pooling), result of which was increased segmentation accuracy. In subsequent work Dinkla et al. (Dinkla et al. 2019) created synthetic CT images from T2-weighted turbo spin echo MRI data. Interestingly, they showed that dental artifacts in CT images, which are problematic for radiotherapy treatment planning, are circumvented in synthetic CT images produced from the MRI data. This is provided measurements have a low sensitivity to magnetic susceptibility effects, such as the case for T2-weighted MRI data. They also stated that radiotherapy treatment dose calculations based on the synthetic CT images were comparable to calculations based on actual CT images.

Wang et al. (Wang et al. 2019) used the standard DCNN U-Net deep learning architecture involving 23 convolutional layers in nasopharyngeal carcinoma. The method was applied in the neck region and the brain. The findings for radiotherapy treatment planning were consistent with those by others, and example images are shown in Figure 3. By going to a larger network, they required a larger dataset for training (23 sets) and testing (10 sets). However, a considerable improvement in synthetic CT image generation was not demonstrated in comparison to other networks where less layers had been implemented and available data for training data was scarce.

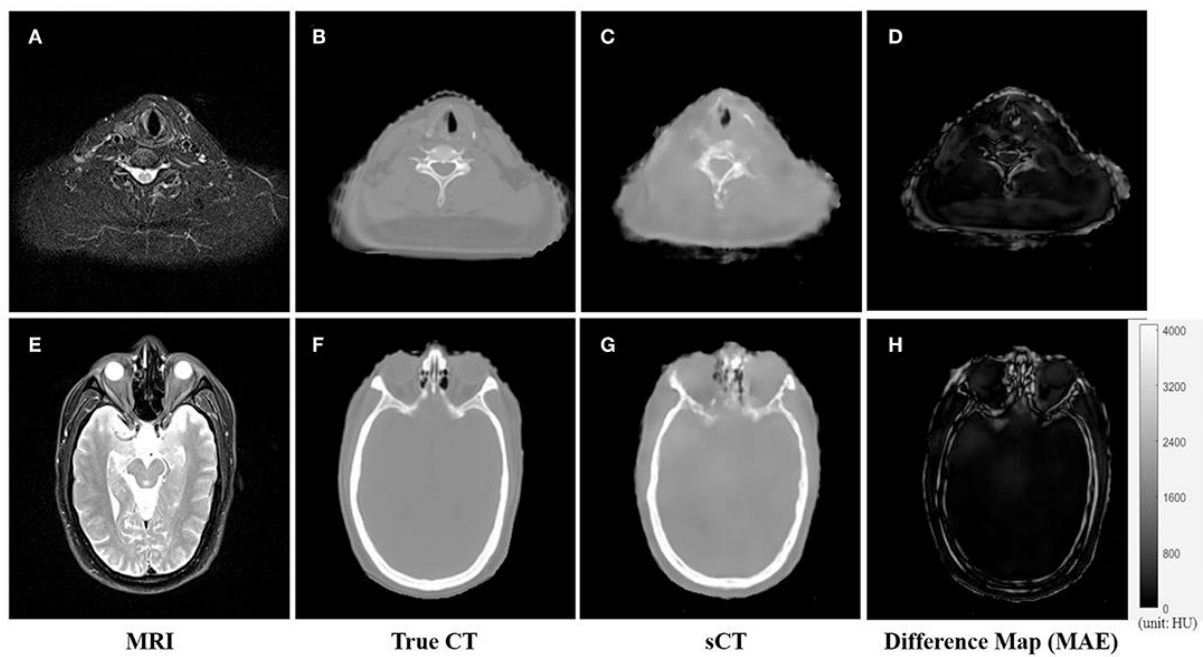

*Figure 3 An example of real T2-weighted MR and real and estimated CT images for neck (top row) and head (bottom row) regions. From left, MR images (input); corresponding CT images (ground-truth); estimated CT; difference map between true and estimated CT images (Wang et al. 2019).*

## 2.2 GAN implementations

Hiasa et al. (Hiasa et al. 2018) synthesised the CT image from MRI and considered the reverse problem as well. A CycleGAN with the addition of gradient consistency loss was proposed to achieve a reliability across anatomical regions. Essentially, the method proposed led to improvements in image intensity estimation over different regions within images. Authors pointed out that image estimation performance differs from the head to joints and muscle. Subsequently, Zhang et al. (Zhang, Yang, and Zheng 2018) found mitigation of geometric distortions in image estimation was not achieved using cycle loss. Therefore, they additionally proposed the use of a shape consistency loss obtained from two segmentation networks. Here, each network segments an image modality into meaningful labels, the result of which is implicit shape constraints for anatomical translation. Semantic labels for training images from both modalities were required to be able to train the network.

Yang et al. (Yang et al. 2018a) also proposed a structure-constrained CycleGAN by adding a structural-consistency loss on top of adversarial and cycle-consistency losses. They used an image modality independent structural feature to map two different modality images into a mutual feature domain, wherein structural consistency was established. Lei et al. (Lei et al. 2019) proposed the use of a dense CycleGAN for dual transformation mapping (i.*e.,* MR → CT & CT → MR simultaneously), in which the dense blocks were replaced with residual blocks in the generator, leading to improved information in estimated images. The use of a distance loss function in addition to a gradient difference loss function was able to compensate for blur and misclassification of image information.

A CycleGAN involving cycle consistent and voxel-wise losses not dependent on paired or unpaired data has been proposed as well for estimating MRI from the CT image (Jin et al. 2019). Peng et al. (Peng et al. 2020) synthesised the CT image from T1-weighted MRI using a cGAN (paired data) and a CycleGAN (unpaired data) and performed training and testing on a nasopharyngeal carcinoma (NPC) patient cohort. The cGAN generator network was a U-Net, and a residual U-Net was used in the CycleGAN implementation. The methods were compared against each other on the same data. The authors reported a better MAE for CycleGAN in comparison with cGAN, see Table 4.

Kazemifar et al. (Kazemifar et al. 2020) generated the synthetic CT image from MRI with the aim of dosimetry evaluation of estimated CT images. They synthesised CT using cGAN with the U-Net generator and the mutual information loss function. The discriminator network involved six convolutional layers followed by five fully connected layers with ReLU as the activation function. The binary cross entropy between two images was used as the loss function. The result showed that the average MAE over all cross-validation groups using cGAN was mostly smaller than results achieved from atlas-based and CNN based methods used in previous studies.

In unrelated research, Rubin and Abulnaga (Rubin and Abulnaga 2019) developed a pix2pix type FCN-cGAN to generate diffusion-weighted MRI from CT perfusion paired data with the aim of improving ischemic stroke lesion segmentation. They considered the case of using CT and the combination of CT and MRI data as input to the network. The combined use of CT and MRI input data to the network provided better ischemic stroke lesion outlines than CT alone.

Yang et al. (Yang, Qian, and Fan 2020) combined VAE and GAN and designed a conditional auto-encoder GAN (CAE-GAN) to synthesise the CT image from multi-contrast MRI data (i.e., fat, water and T2 values). Example image estimates are shown in Figure 4. The proposed method resulted in lower MAE and high correlation coefficients compared with the FCM machine learning clustering (Bezdek, Ehrlich, and Full 1984) and the FCN deep learning (Nie et al. 2016b) methods.

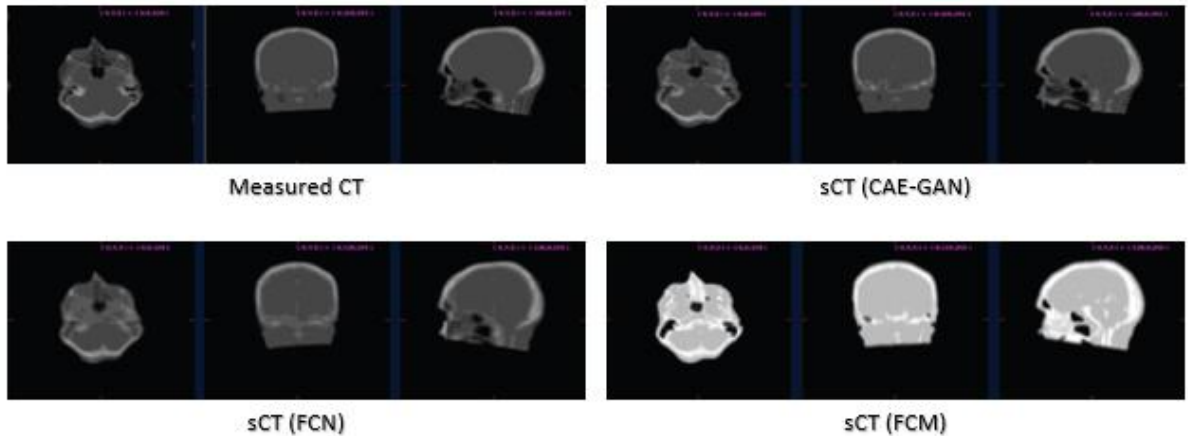

*Figure 4 An example of CAE-GAN performance and its comparison with FCN and FCM based methods to synthesise CT. Images reproduced from (Yang, Qian, and Fan 2020).*

Boni et al. (Brou Boni et al. 2020) synthesised the CT image from MRI in the pelvic region using a previous network developed for non-medical image synthesis (Wang, Liu, et al. 2018). They introduced two auto-encoder generator sub-networks and called it the coarse-to-fine generator, in which one was responsible for translating low-resolution images, and the action of the other was to refine images into a high-resolution output. Also, a multi-scale discriminator involving three networks was developed in which each function acted at different image scales. For optimisation of the proposed method, the feature matching loss was incorporated into the networks to steady the training by forcing the generator to estimate images with natural statistics at various scales. Finally, cross-entropy cGAN losses in the pix2pixHD network were replaced with Least Square GAN (LSGAN), a key element of which is the optimisation of the gradient vanish during generator updates. Estimated images are shown in Figure 5. The comparison between estimated and real CT images acquired from patients using Pix2pixHD network outperformed vanilla pix2pix (Isola et al. 2017) based on the MAE results (48.5 ± 6 vs 62.0±12 HU).

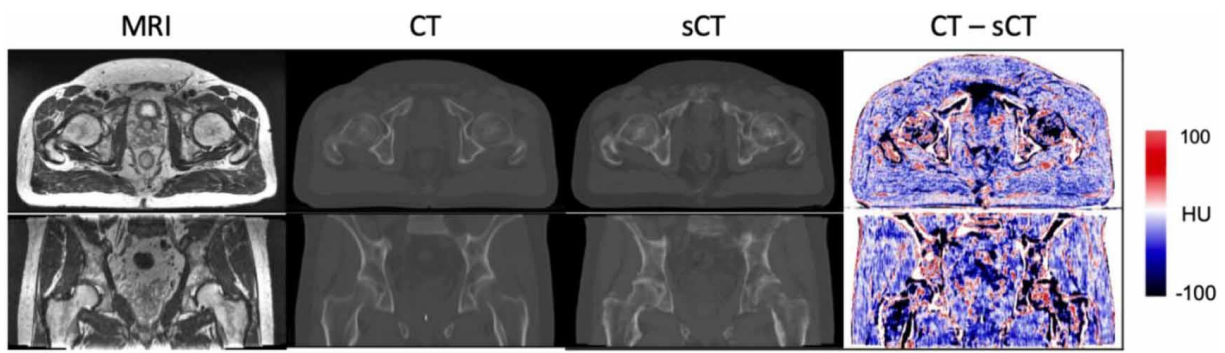

*Figure 5 An example of sCT estimated from MRI. The top row and bottom rows represent axial and frontal planes respectively from (Brou Boni et al. 2020).*

Fetty et al. (Fetty et al. 2020) generated the synthetic CT image from MRI based on a more generalised network, compatible with MRI acquired at different field strengths. They changed the cGAN generator to an ensemble model of four different network outputs (i.e., SE-ResNet, DenseNet, U-Net and Embedded Net). On the same data the ensemble model was compared with networks using only SE-ResNet, DenseNet, U-Net or Embedded Net as the generator. In each case the ensemble model estimated CT image was of better quality than those produced by any of the other networks.

Tie et al. (Tie et al. 2020) proposed a novel GAN method for estimating the CT image from MRI using multi-channel multi-path generative adversarial network (MCMP-GAN). In this method a 5-level residual U-Net was implemented as the generator where each channel (i.e. MRI corresponding to precontrast T1-weighted, postcontrast T1-weighted with fat-saturation, and T2-weighted) was responsible for extracting features from individual MRI contrasts. The discriminator was a CNN with five convolutional layers. They compared the proposed method using MAE, PSNR and DSC with multi-channel single-path and single-channel single-path GANs. The results of MAE, PSNR and DSC showed the MCMP-GAN to outperform both multi-channel single-path and single-channel single-path GANs.

Liu et al. (Liu et al. 2021) addressed the benefit of dual energy CT (DECT), a method known to have benefits in making CT-dose calculations. They proposed a CycleGAN model to generate DECT images (i.e. high energy and low-energy CT) from MRI with the addition of a generator for inverse mapping (i.e. generating MRI from CT). The proposed CycleGAN, called "label GAN", was able to discriminate real and synthetic DECT as well as distinguish between high-energy and low-energy CT images from DECT.

### 2.3 How do the networks compare?

The CNN and GAN based studies have aimed to solve the same image estimation problem, often for different anatomical sections using different data. Whilst it is difficult to directly compare the pros and cons of each method, some general observations can be made. The autoencoder takes images as inputs and extracts features from the images. The use of U-Net, a state-of-the-art autoencoder, appears to provide better image synthesis than when such a framework is not used. Additionally, dimensionality reduction from 3D volumes to 2D slices to 2D patches also has some key advantages in terms of training and testing of CNNs and GANs, i.e. incremental improvements in quality of estimated images. The GAN being an extension of a CNN through the incorporation of a discriminator network provides the additional benefit of even better-quality image estimates. The CycleGAN has the extra bonus of not having to co-register cross-modality images prior to network training. Findings suggest the best cross-modality image estimates are ones where a CycleGAN is used in the absence of co-registered input images.

## 3. PET IMAGE ESTIMATION FROM MRI

### 3.1 CNN implementations

Li et al. (Li et al. 2014b) were interested in predicting missing slices in a 3D medical imaging dataset from a full 3D dataset acquired using a different imaging modality. Essentially, the problem defined by them was one involving a CNN for cross-modality image estimation. To achieve this goal, a CNN was trained using 3D MRI patches of size 15 x 15 x 15 voxels resulting in a 3D PET patch of 3 x 3 x 3 pixels. Notably, the MRI data had five times higher resolution, hence the low-resolution PET patch estimation. The CNN here appeared to work by creating an optimised network which learnt the anatomical structures within the human brain based on those captured in MRI data. Then, missing information can be predicted based on expected anatomy, since tracer uptake in PET is preferentially distributed based on anatomical variations. The CNN model did not incorporate pooling, and as such information loss through resolution reduction was not expected. The 3D CNN was trained using 50,000 patches from a total of 199 subjects who had corresponding MRI and PET data. Evaluation of image estimation involved binary-class classification to differentiate between healthy controls, and Alzheimer's disease and mild cognitive impairment patients. The 3D-CNN PET image estimate was used in the classification and was able to match ground truth, and those produced by the K-nearest neighbour and Zero methods (Yuan et al. 2012) were not able to achieve a similar level of disease classification.

## 3.2 GAN implementations

Pan et al. (Pan et al. 2018) developed a 3D CycleGAN for estimating PET from MRI using registered datasets. They evaluated PSNR and classification accuracy based on the real and estimated PET images. The authors suggested the method to be sufficiently reliable for clinical use. A comparison of estimated and real images is provided in Figure 6.

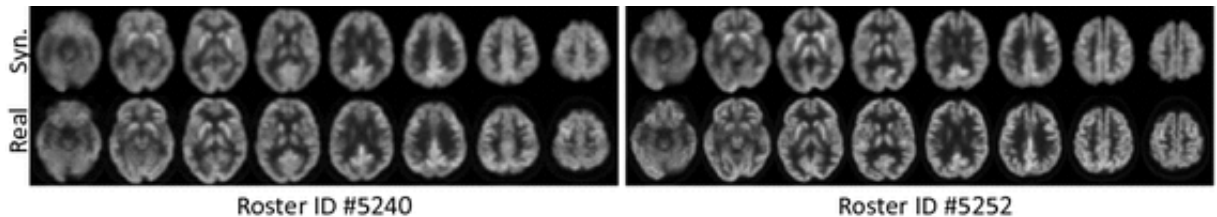

*Figure 6 Comparison of synthetic and real images for two subjects reproduced from (Pan et al. 2018).*

In a multiple sclerosis study on myelination, Wei et al.(Wei et al. 2018) also estimated PET from paired MRI data using a combination of two 3D cGANs - called the Sketcher-Refiner GAN. The Sketcher generates a preliminary image capturing anatomical and physiological aspects of the brain, whilst the Refiner enhances the preliminary image to better reflect myelin content. The method outperformed the CNN proposed by Li et al. (Li et al. 2014b) as well as a single cGAN where a Refiner network is not included.

# 4. PET IMAGE SYNTHESIS TO BOOST RESOLUTION AND SNR AIDED BY MRI

Many researchers have focused on denoising PET images using deep learning methods to boost signal-to-noise ratio, for example Gong et al. (Gong et al. 2020) Particularly, the interest has been on using low dose PET data acquisitions and predicting the high dose equivalent images. PET is also a low resolution imaging modality, and the use of MRI data in deep learning approaches has been proposed for enhancing PET image resolution (Turkheimer et al. 2008).

## 4.1 CNN implementations for converting low dose PET to high dose images

In 2017, Xiang et al. (Xiang et al. 2017) worked on estimating high quality standard-dose PET images from a low quality low-dose PET and considered the incorporation of T1-weighted MRI data. A CNN network designed for this study contained three steps. The first CNN step was to concatenate the T1-weighted MRI and low dose PET patches, followed by four convolutional layers to generate a standard dose PET patch. In the second step, the concatenated input patches were added to the last feature map from the first step, becoming inputs to another four convolutional layers. The same process was repeated, and the final estimated image was the output of the third, four-layer network. Hence, the network had a total of 12 layers. The authors emphasised the omission of pooling, since it results in dimensionality reduction of the feature map, a property not desirable for pixel-wise image quality enhancement. Sample size limitations (16 subjects) were overcome by replacing the slice-by-slice input with a patch based volumetric input to make available more training data for the network. The computed training and estimation times using this CNN network was 4.2 hours and 2s, respectively, compared with 2.9 hours and 16 minutes for a sparse learning method (i.e. multilevel canonical correlation analysis for standard dose PET image estimation). These two approaches performed comparatively as measured using PSNR and NMSE (PSNR – 24.76 vs 24.67; NMSE – 0.0206 vs 0.0210). Based on a comparison between the proposed CNN and a more conventional 12-layer network, the new CNN outperformed the conventional approach in terms of PSNR (24.76 vs 23.98) and NMSE (0.0206 vs 0.0247). The use of T1-weighted MRI data in addition to low-dose PET images beat the case when low dose PET was the only input data for high dose PET image estimation (PSNR - 23.85 vs 23.11; NMSE - 0.0254 vs 0.0299). Example images are shown in Figure 7.

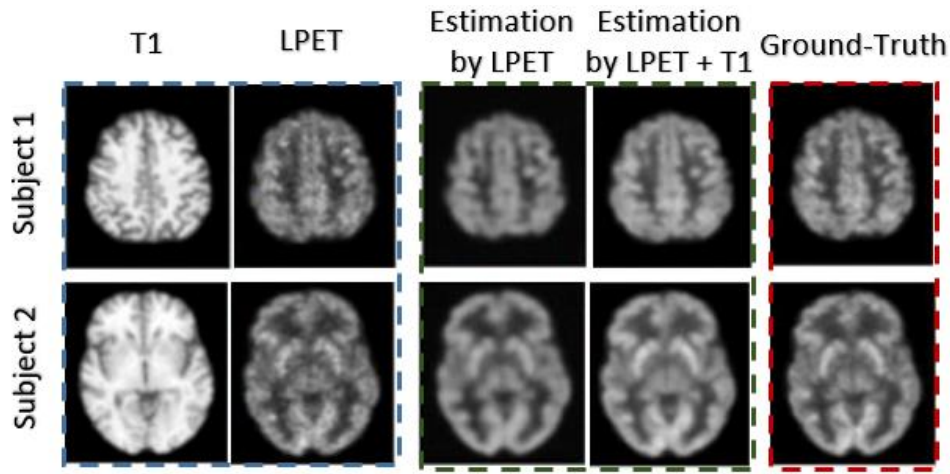

*Figure 7 Example of two subjects and comparison between their estimated high-dose PET from low-dose PET, estimated high-dose PET from low-dose PET and MRI, and ground-truth. Adapted and modified from (Xiang et al. 2017).*

## 4.2 CNN implementations for PET resolution enhancement

Costa-Luis et al. (Costa-Luis and Reader 2017) proposed a post-processing step to improve PET image resolution with the aid of MRI data. A 3-layer CNN was trained using simulated T1-weighted MRI and PET data. In comparison with the state-of-the-art maximum likelihood expectation maximisation (MLEM) image reconstruction method, the CNN approach could suppress noise, ringing and partial volume artefacts alongside an 80% improvement in the normalised root mean squared error (NRMSE) between estimated and expected images.

Inspired by Han's study (Han 2017) on CT image synthesis from MRI, Liu et al. (Liu and Qi 2019) proposed two different DNNs adopting a U-Net architecture for PET image signal-to-noise ratio improvement. The network consisted of 25 convolutional layers with skip connections between layers of identical input dimensions in the encoding and decoding parts of the U-Net. Max pooling was introduced to accelerate training by down sampling feature maps. A model using three U-Nets was proposed as well for the same purpose. The primary difference between the 1-U-Net and 3-U-Net CNNs was how the inputs were handled. In the 1-U-Net, the input consisted of two channels, i.e. the low signal-to-noise ratio PET and T1-weighted MRI data together. In the 3-U-Net implementation, the PET and MRI data were fed to independent 1-U-Nets, before the output was concatenated into another U-Net, resulting in the 3-U-Net architecture. In both cases the target output used for training was the MLEM PET reconstruction. The training data included 21 T1-weighted MRI and noise-free PET digital brain phantoms generated from real subjects. Poisson noise was added to the sinogram to create low signal-to-noise ratio PET images. The study considered training with and without MRI data. In comparison with 1-U-Net trained only using low signal-to-noise ratio PET images, the additional use of MRI data led to a decrease in MSE by 31.3% and 34% for 1-U-Net and 3-U-Net, respectively. Contrast improvements of 2.7 and 1.4 fold for 1-U-Net and 3-U-Net were also reported.

Recently, Song (Song et al. 2020a) proposed a deep learning method for PET image resolution enhancement assisted by high resolution MRI data. In the CNN architecture they introduced the use of multichannel patch-based input as separate convolutions matched with patch spatial locations. The multiple channels were low-resolution PET and high-resolution T1-weighted MRI, supplemented with radial and axial coordinates for patch location. The image generation task was performed using three different networks, i.e. shallow CNN (3 layers), deep CNN (12 layers), and very deep CNN (20 layers). Four alternatives as input data were considered, namely PET patches, corresponding PET and MRI patches, PET patches with radial and axial location, and corresponding PET and MRI patches with radial and axial location. Contrast-to-noise ratio, PSNR and SSIM results showed that there was a significant improvement in high resolution PET image estimation with the use of MRI data, as demonstrated by others. Patch location information provided limited benefits. The PSNR and SSIM metrics were found to increase from the shallow to deep CNN structures. Qualitative differences between the shallow and very deep CNN results were hard to discern when MRI was used as input to the network.

### 4.3 GAN implementations for PET resolution enhancement

Wang et al. (Wang, Zhou, et al. 2018) proposed a cGAN named Locality-adaptive GAN (LA-GAN) to improve PET image resolution using paired data. The proposed LA-GAN consisted of three networks; the locality adaptive network for adjusting spatial differences between MRI and PET images, and generator and discriminator networks. Subsequently, Song et al. (Song et al. 2020b) considered the use of a CycleGAN with unpaired data. They used the generator from the CNN trained on paired data to reduce the CycleGAN training time. The actual implementation of the CycleGAN involved one generator and two discriminators. Their CycleGAN achieved the best results to date when compared with other state-of-the-art methods trained using unpaired data. Estimated super resolution images using different methods are shown in Figure 8.

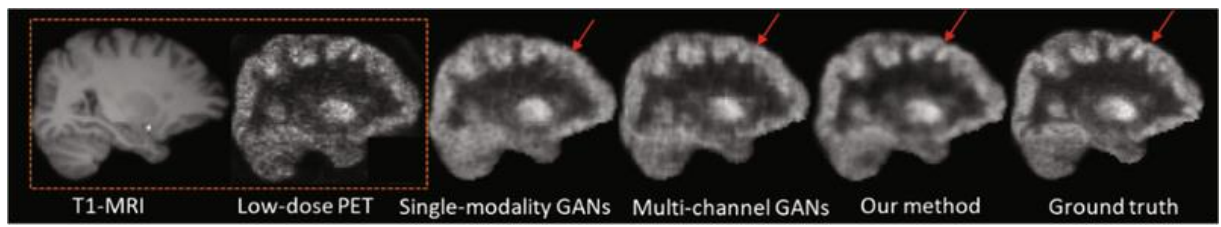

*Figure 8 Comparison of estimated super resolution image for one subject using single-modality (Low dose PET only, Multi-modality (low dose PET and MRI) and LA-GAN (low dose PET and MRI). Reproduced from (Wang, Zhou, et al. 2018).*

## 5. FINAL REMARKS

Cross-modality image estimation methods involving MRI data have received increasing attention in the field of medical imaging. We reviewed recent advances in deep learning, i.e. focusing on CNNs and GANs, for estimating one imaging modality image from another modality. We focused on the neural network structures and how image estimations were evaluated. An ideal cross-modality image estimation framework involves 3D imaging data for training and estimation, which may have to be compromised under certain circumstances. However, a key requirement of CNN and GAN training is large datasets from which the machine can learn, a prerequisite which may be plausible in healthy participants but unattainable for patient cohorts. Studies to date have proposed techniques to overcome existing challenges.

### 5.1 Is cross-modality image estimation valid?

Cross-modality image estimation based on neural networks involves taking one biomedical image and converting it to an image otherwise generated using a different instrument. The action of performing this task using a neural network is to look for patterns in the image and associate the identified features with patterns expected in the target image. Based on this, image estimation boils down to a pattern recognition and association problem. But, cross-modality image estimation, in principle, should be the estimation of the correspondence between the physics of different imaging modalities. For example, estimating $^{18}$FDG-PET from MRI suggests that information across these modalities is consistent. Whilst this may be true for this PET agent, since tracer uptake localises to all soft tissue classes, and MRI is a soft tissue imaging modality, it may not be the case for a tracer specific only to cancer. For this reason, in the future consideration should be made towards the underlying physics of each modality when developing neural networks for cross-modality image estimation. There has been increasing interest in physics inspired neural networks, (Willard et al. 2020) potentially providing a mechanism by which existing limitations can be overcome. Here, the neural network may be developed to predict correspondence between tissue classes, such as grey and white matter, for example, instead of associating patterns between different imaging modalities. In this way smaller neural networks become plausible as well.

## 5.2 Is dimensionality reduction useful?

Three specific ways can be used to overcome existing challenges associated with a lack of training data. First, divide 3D volumetric images into smaller patches covering the entire 3D volume. This reduces the dimension of the image input to the network, the result of which is reduced CNN complexity and computational resource requirements decrease as well. With similar benefits, the second approach is to create 2D slices from 3D image volumes, which can later be stacked to recover the 3D image. The number of patches and their size for optimal training of the CNN/GAN remains an open question. Lastly, like patches, pooling reduces the number of CNN/GAN parameters, which poses another solution for reduction in computational overheads. A negative consequence of pooling is information loss through a reduction in spatial resolution of image features, an undesirable property for cross-modality image estimation.

## 5.3 Is this the end or just the beginning?

Future work on image estimation should consider the improvement of the training accuracy by optimising patch characteristics, investigating the impact of the number of datasets used for training, and accelerate training and prediction times by creating more efficient neural networks. People may also wish to consider the application of existing networks to anatomical regions outside of those investigated to date. Moreover, the true power of cross-modality image estimation may only be realised when the applicability of machine learning frameworks have been demonstrated for diseases and disorders where medical imaging plays a crucial role in diagnosis, treatment planning and monitoring.

In summary, cross-modality image estimation involving MRI as one of the modalities has been investigated widely using CNNs and GANs. The various studies reviewed used different cross-modality image types, inconsistent training and testing dataset sizes, and images generally consisted of different anatomical locations. Nonetheless, it does appear that GANs can outperform CNNs in cross-modality image estimation. The CycleGAN, involving two autoencoders with unpaired data as inputs for the generator and PatchGAN as the discriminator, was shown to outperform other types of GANs involving paired cross-modality input data. The CycleGAN does come at a cost of increased training effort, an unfortunate consequence of requiring extra training datasets and computational resources for cross-modality image estimation generalisability.

*Table 3. CNN architecture summary and parameters required for training, + represents MSE, * represents MAE, "1" and "2" represents 1-norm (L1) and 2-norm (L2) loss function. N/A: Not Applicable, NS: Not Stated, ND: Not disclosed by author.*

| Author-year | Image estimation | Architecture | Remarks | Sample size (Training-Testing) | Patch size | Number of patches (from each volume) | Number of Convolutional layers | Pooling layers Y/N? | Error (HU) Mean ± standard deviation |
|---|---|---|---|---|---|---|---|---|---|
| (Li et al. 2014b) | T1-weighted MR: PET | 3D CNN | Brain | 199 | 3 x 3 x 3 | 50,000 | 2 | N | NS |
| (Nie et al. 2016b) | T2-weighted MR: sCT | 3D FCN | Pelvic | 22 | MRI: 32 x 32 x 16 CT: 24 x 24 x 12 | 6000 | 2 | N | 42.4 ± 5.1*1 |
| (Costa-Luis and Reader 2021) | (Low count PET + T1-Weighted MR): full count PET | 3D CNN | Brain | 20 | N/A | N/A | 3 | N | +2 |
| (Xiang et al. 2017) | (Low-dose-PET + T1-Weighted MR): standard-dose-PET | 2D CNN | Brain | 16 | Step 1: 27 x 27 Step 2: 19 x 19 Step 3: 11 x 11 | 30,000 | 4 | N | 0.0206+2 |
| (Roy, Butman, and Pham 2017) | (UTE) dual-echo MR: sCT | 3D CNN (Inception blocks adopted) | Brain | 6 | 25 x 25 x 5 | 500,000 | 6 | Y | PSNR:21.92 +2 |
| (Han 2017) | T1-Weighted MR: sCT | 2D DCNN, (U-Net adopted) | Brain | 18 | N/A | N/A | 27 | Y | 84.8 ± 17.3*1 |
| (Liu et al. 2017) | MR: CT | 2D Auto-encoder network | Brain | 30-10 | N/A | N/A | 26 | Y | Multiclass cross-entropy loss |
| (Fu et al. 2019) | T1-Weighted MR: sCT | 2D & 3D CNN (U-Net adopted) | Pelvic | 20 | N/A | N/A | 27 | Y | 2D: 40.5 ± 5.4*1  3D: 37.6 ± 5.1*1 |
| (Arabi et al. 2018) | T2-Weighted MR: sCT | 2D DCNN (U-Net adopted) | Pelvic | 39 | N/A | N/A | 27 | NS | 18.4 ± 6.6*1 (bladder) 78.3 ± 69.2*1 (rectum) 32.7 ± 7.9*1 (body) 119.9 ± 22.6*1 (bone) |
| (Chen, Qin, et al. 2018) | T2-Weighted MR: sCT | 2D DCNN (U-Net adopted) | Pelvic | 36-15 | N/A | N/A | 23 | Y | 29.96 ± 4.87*1 (within body) |
| (Leynes et al. 2018) | ZTE MR + Dixon(fat and water) MR: sCT | 3D DCNN (U-Net adopted) | Pelvic | 10-16 | 32 x 32 x 16 | 60,000 | 13 | NS | −36 ± 130+2 (Dixon pseudo-CT) −12 ± 78+2 (ZeDD CT) |
| (Xiang et al. 2018) | T1-Weighted MR: sCT | 3D DECNN (Embedding blocks) | Brain | 16 | 64 × 64 x 3 | Brain: 600,000 | 3K+5 (K indicates the number of embedding blocks) 11 | N | 85.4 ± 9.24*1 |
|  |  |  | Pelvic | 22 |  | Prostate:100,000 | 17 |  | 33.5 ± 0.8*1 |
| (Dinkla et al. 2018) | T1-Weighted MR: sCT | 2.5D dilated CNN | Brain | 52 | N/A | N/A | 10 | N | 67 ± 11*1 |
| (Dinkla et al. 2019) | T2-Weighted MR: sCT | 3D (U-Net adopted) | Head and Neck | 22 | 48 × 48 × 48 | 8 | 14 | NS | 75 ± 9*1 |
| (Wang et al. 2019) | T2-Weighted MR: sCT | 2D (U-Net adopted) | Head and Neck | 23-10 | N/A | N/A | 23 | Y | 97 ± 13*1 (soft tissue), 131 ± 24*1 (overall region), 357 ± 44*1 (bone) |
| (Torrado-Carvajal et al. 2019) | Dixon -Vibe MR: sCT | 2D (U-Net adopted) | Pelvic | 15-4 | N/A | N/A | 23 | Y | *1 |

| Author-year | Image estimation | Architecture | Anatomy | Training data | Validation data | Testing data | Epochs | Paired cross-modal data (Y/N) | Error (HU) Mean ± standard deviation |
|---|---|---|---|---|---|---|---|---|---|
| (Florkow et al. 2020) | T1-weighted multi-echo gradient-echo acquisition: sCT | 4D (3 spatial dimensions and 1 channel dimension) (U-Net adopted) | Pelvic | 17 canines, 23 humans | C × 24 × 24 × 24 ("c" is number of channels) | NS | NS | Y | 33-40[*1] (human) 35-47[*1] (canines) |
| (Gupta et al. 2019) | T1-Weighted MR: sCT | 2D DCNN, (U-Net adopted) | Brain | 47-13 | N/A | N/A | NS | NS | 17.6 ± 3.4[*1] (soft tissue) [*1 & +2] |
| (Spadea et al. 2019) | T1-Weighted MR: sCT | 2D DCNN, (U-Net adopted) | Brain | 15 | N/A | N/A | 27 | Y | 54 ± 7[*1] (within the head) |
| (Chen, Gong, et al. 2018) | (Low dose 18F-florbetaben PET+ T1-W, T2-W & T2-FLAIR MR): high SNR PET | 2D DCNN, (U-Net adopted) | Brain | 32-8 | N/A | N/A | 22 | Y | 0.143725[+1] |
| (Liu and Qi 2019) | (Low SNR PET + T1-Weighted ME): high SNR PET | 3U-Net (U-Net adopted) | Brain | 20 | N/A | N/A | 26 | Y | [+2] |
| (Liu et al. 2020) | T1-Weighted MR Dixon: sCT | 2.5D (U-Net adopted) | Abdominal | 46 | N/A | N/A | 23 | Y | 24.10[*1] (liver), 28.62[*1] (spleen), 47.05[*1] (kidneys), 29.79[*1] (spinal cord), 105.68[*1] (lungs), 110.09[*1] (vertebral bodies) |
| (Bahrami et al. 2020) | T2-Weighted MR: sCT | 2D eCNN | Pelvic | 12 | N/A | N/A | 52 | Y | 30.0 ± 10.4[*1] |
| (Song et al. 2020a) | (Low resolution PET + T1-Weighted MR): super-resolution PET | 2D DCNN | Brain | 15 | NS | NS | Shallow CNN: 3 Deep CNN: 12 Very deep CNN: 20 | N | [*1] NS |
| (Victor et al. 2021) | T1 & T2-Weighted MR: sCT | 4D (3 spatial dimensions and 1 channel dimension) (U-Net adopted) | Lumbar Spine | 3 | ND | ND | ND | ND | ND |

Table 4. Provided is a summary of cross-modality image estimation using GANs. Abbreviations: High-energy CT: HECT low-energy CT: LECT

| Author-year | Image estimation | Architecture | Remarks | Paired cross-modal data (Y/N) | Error (HU) Mean ± standard deviation |
|---|---|---|---|---|---|
| (Wolterink et al. 2017) | MR: CT | CycleGAN (Residual blocks: generator; PatchGAN: discriminator) | Brain | N | MAE 73.7 ± 2.3 |
| | MR: CT | Cascade GAN | Brain & Pelvic | Y | Brain: Mean MAE 92.5 ± 13.9 |

| Reference | Modality | Network | Site | Paired | Results |
|---|---|---|---|---|---|
| (Nie et al. 2018) | | (FCN: generator; CNN: discriminator) | | | Mean PSNR 27.6 ± 1.3 |
| | | | | | Pelvic: Mean MAE 39.0 ± 4.6<br>Mean PSNR 34.1 ± 1.0 |
| (Emami et al. 2018) | MR: CT | cGAN (ResNet: generator; 5 conv-layer CNN: discriminator) | Brain | Y | Mean PSNR 26.6 ± 1.2,<br>mean SSIM 0.83 ± 0.03<br>MAE 89.3 ± 10.3 (Across the entire FOV) |
| (Maspero et al. 2018) | MR: CT | cGAN (Pix2pix) (U-Net: generator; PatchGAN: discriminator) | Pelvic | Y | MAE 61 ± 9<br>ME 2 ± 8 |
| (Hiasa et al. 2018) | MR: CT | CycleGAN (Residual blocks: generator; PatchGAN: discriminator) | Musculoskeletal | N | MAE 29.781 ± 1.777 |
| (Huo et al. 2018) | MR: CT | CycleGAN (EssNet) (Residual blocks: generator; PatchGAN: discriminator) | Abdomen | N | Median DSC 0.9188 |
| (Lei et al. 2019) | MR: CT | CycleGAN (Dense blocks: generator) | Brain & Pelvic | Y | Brain images: mean MAE 55.7, mean PSNR 26.6<br>Pelvic images: mean MAE 50.8, mean PSNR 24.5 |
| (Peng et al. 2020) | MR: CT | cGAN (U-Net: generator; 6 conv-layer CNN: discriminator) | Head-neck-spine | Y | MAE: 69.67 ± 9.27 |
| | | CycleGAN (Residual U-Net: generator; 6 conv-layer CNN: discriminator) | | N | MAE: 100.62 ± 7.39 |
| (Yang, Qian, and | MR: CT | CAE-GAN (VAE + GAN) | Brain | Y | MAE: 81.25 |
| Fan 2020) | | | | | |
| (Yang et al. 2018b) | MR: CT | CycleGAN (Residual blocks: generator; PatchGAN: discriminator) | Brain | N | The value of MAE and PSNR are Not Stated |
| (Brou Boni et al. 2020) | MR: CT | cGAN (Pix2pixHD) (ResNet: generator; PatchGAN: discriminator) | Pelvic | Y | MAE 48.5 ± 6 |
| (Tie et al. 2020) | MR: CT | cGAN (MCMP-GAN) (Residual U-Net: generator; 5 conv-layer CNN: discriminator) | Head (nasopharyngeal carcinoma) | Y | MAE 75.7 ± 14.6 |
| (Fetty et al. 2020) | MR: CT | cGAN (Pix2pix) (ensemble generator consists of SE-ResNet, DenseNet, U-Net, and Embedded Net; PatchGAN: discriminator) | Pelvic | Y | MAE 41.2 ± 3.7 (0.35T Siemens Magnetom C!) MAE 52.0 ± 5.5 (1.5T Siemens) MAE 43.7 ± 6.2 (3T GE Discovery) MAE 48.2 ± 4.9 (3T GE Signa) |
| (Kazemifar et al. 2020) | MR: CT | cGAN (U-Net: generator; 6 conv + 5 fully connected layers CNN: discriminator) | Brain | Y | MAE 47.2 ± 11.0 |
| (Tang et al. 2021) | MR: CT | cGAN (pix2pix) (U-Net: generator; 5 conv-layer CNN: discriminator) | Brain | Y | MAE 60.52 ± 13.32 |
| (Liu et al. 2021) | MR: CT | CycleGAN (label GAN) (encoder-decoder FCN: generator; CNN: discriminator) | Head and neck | Y | MAE LECT 79.98 ± 18.11 MAE HECT 80.15 ± 16.27 |

| Study | Modality | Method | Region | Paired | Results |
|---|---|---|---|---|---|
| (Jin et al. 2019) | CT: MR | CycleGAN | Brain | N | MAE 19.36 ± 2.73, mean PSNR 65.35 ± 0.86, mean SSIM 0.25 ± 0.02 |
| (Jiang et al. 2019) | CT: MR | CycleGAN (generator network was adopted from (Radford, Metz, and Chintala 2015)); PatchGAN: discriminator) | Lung | N | DSC 0.75 ± 0.12 |
| | | UNIT | | | DSC 0.63 ± 0.22 |
| (Rubin and Abulnaga 2019) | CT: MR | cGAN (Pix2pix) (FCN: generator; PatchGAN: discriminator) | Brain | Y | Dice 0.54 ± 0.23 Hausdorff Distance 27.88 ± 21.00 |
| (Chartsias et al. 2017) | CT: MR | CycleGAN | Heart | N | Average dice score for real and synthetic data 0.704 |
| (Zhang, Yang, and Zheng 2018) | MR ⇌ CT | CycleGAN* (U-Net: generator; PatchGAN: discriminator) | Heart | N | Dice score estimated CT 74.4% Estimated MR 73.2% |
| (Huo et al. 2019) | MR: CT | CycleGAN( SynSeg-Net) (Residual blocks: generator; PatchGAN: discriminator) | Abdomen | N | Median DSC 0.919 |
| | CT: MR | | Brain | | Median DSC 0.966 |
| (Wei et al. 2018, 2019) | MR: PET | cGAN (Sketcher-Refiner GANs) (U-Net: generator; 4 conv-layer CNN: discriminator) | Brain | Y | MSE 0.0083 ± 0.0037 PSNR 30.044 ± 1.095 |
| (Pan et al. 2018) | MR: PET | CycleGAN (Encoding-transferring-decoding CNN: generator; 5 conv-layer | Brain | Y | Mean ± Std PSNR 24.49 ± 3.46 |

| Reference | Modality | Method | Region | Clinical application (Y/N) | Results |
|---|---|---|---|---|---|
| | | CNN: discriminator) | | | |
| (Choi and Lee 2018) | PET: MR | cGAN (Pix2pix) | Brain | Y | SSIM 0.91 ± 0.04 MAE 0.04 ± 0.03 |
| (Wang, Zhou, et al. 2018) | (Low resolution PET + MR): super-resolution PET | cGAN (LA-GANs) | Brain | Y | NC subjects 24.58 ± 1.78 MCI subjects 25.15 ± 1.97 |
| (Song et al. 2020b) | (Low resolution PET + MR): super-resolution PET | CycleGAN | Brain | N | SSIM 0.941 PSNR 35.57 |
| **Cross-modality image estimation excluding MR** | | | | | |
| (Bi et al. 2017) | CT: PET | cGAN (pix2pix-M-GAN) (U-Net: generator; 5 conv-layer CNN: discriminator) | Lung | Y | MAE 4.60 PSNR: 28.06 |
| (Ben-Cohen et al. 2019) | CT: PET | FCN-cGAN (U-Net: generator; 4 conv-layer CNN: discriminator) | Liver | Y | MAE 0.72 ± 0.35, mean PSNR 30.22 ± 2.42 |
| (Armanious, Jiang, et al. 2020) | PET: CT | cGAN (MedGAN) (U-Net: generator; modified PatchGAN: discriminator) | Brain | Y | MSE 264.6 |
| (Armanious, Hepp, et al. 2020) | PET: CT | cGAN (Similar to MedGAN) | Whole body | Y | Mean difference -1.5 ± 47.3 |
| (Vitale et al. 2020) | CT: US | CycleGAN (Both U-Net AND ResNet used individually as generators and results were compared) | Abdomen | N | NS |

# 6. Appendices

## APPENDIX A: Material and method

### A.1 Inclusion criteria

To find all the relative studies for our review, we searched based on the following alternative words via UQ library in both PubMed, Scopus, and Google Scholar databases.

*(structural imag OR molecular imag OR clinical imag OR medical image) AND (image OR imaging) AND (cross-modal OR cross modal OR multimodal OR multi-modal OR intermodality OR intermodal) AND (image estimation OR image prediction OR image generation OR reconstruction OR translation) AND (pseudo OR synthetic OR synthesis OR artificial) AND ( MRI OR CT OR PET OR SPECT )AND (artificial intelligence OR AI OR machine learning OR ML OR deep learning OR convolutional neural network OR neural network OR CNN OR artificial neural network OR ANN OR Adversarial Network OR GAN) NOT radiomics NOT outcome prediction NOT intramodality)*

### A.2 Exclusion criteria

The following three exclusion criteria for article selection were applied: i) non-medical image estimation, ii) non-CNN and non-GAN based image estimation methods, for instance random forest, iii) hybrid PET-MRI (i.e., only one input) scan as an input and other PET (for instance high-dose PET) scan as a ground truth.

### A.3 Literature search

The study reviewed the estimation of one imaging modality image from another using CNNs and GANs, both promising deep learning methods in computer vision for pattern discovery. Reviews in 2019 and 2020 have been published on GANs, primarily focusing of image synthesis including 21 works involving cross-modality image estimation (Sorin et al. 2020; Yi, Walia, and Babyn 2019). The contributions in this space greatly increased over the past few years. To date, the following number of papers involved cross-modality image estimation using GANs: 21 CT to MRI or vice versa; 3 MRI to PET or vice versa; 4 CT to PET or vice versa; and 1 CT to ultrasound. Whilst this review is primarily aimed at the MRI community, the five non-MRI papers have been included for completeness, refer to Table 5.

*Table 5. List of articles included based in the inclusion criteria in the review.*

| Author- year | Title | Journal |
|---|---|---|
| **Convolutional Neural Networks** | | |
| Li et al., 2014(Li et al. 2014b) | Deep Learning Based Imaging Data Completion for Improved Brain Disease Diagnosis. | Paper presented at the Medical Image Computing and Computer-Assisted Intervention. |
| Nie et al., 2016(Nie et al. 2016b) | Estimating CT Image from MRI Data Using 3D Fully Convolutional Networks. | Paper presented at the Deep Learning and Data Labelling for Medical Applications. |
| Costa-Luis et al., 2017(Costa-Luis and Reader 2017) | Deep Learning for Suppression of Resolution-Recovery Artefacts in MLEM PET Image Reconstruction. | Paper presented at the 2017 IEEE Nuclear Science Symposium and Medical Imaging Conference (NSS/MIC). |

| Xiang et al., 2017(Xiang et al. 2017) | Deep auto-context convolutional neural networks for standard-dose PET image estimation from low-dose PET/MRI. | Neurocomputing |
|---|---|---|
| Roy et al., 2017(Roy, Butman, and Pham 2017) | Synthesizing CT from Ultrashort Echo-Time MR Images via Convolutional Neural Networks. | Paper presented at the Simulation and Synthesis in Medical Imaging, Cham. |
| Han, 2017(Han 2017) | MR-based synthetic CT generation using a deep convolutional neural network method. | Medical Physics |
| Liu et al., 2017(Liu et al. 2017) | Deep Learning MR Imaging–based Attenuation Correction for PET/MR Imaging. | Radiology |
| Fu et al., 2019(Fu et al. 2019) | Deep learning approaches using 2D and 3D convolutional neural networks for generating male pelvic synthetic computed tomography from magnetic resonance imaging. | Medical Physics |
| Arabi et al., 2018(Arabi et al. 2018) | Comparative study of algorithms for synthetic CT generation from MRI: Consequences for MRI-guided radiation planning in the pelvic region. | Medical Physics |
| Chen et al., 2018(Chen, Qin, et al. 2018) | Technical Note: U-Net-generated synthetic CT images for magnetic resonance imaging-only prostate intensity-modulated radiation therapy treatment planning. | Medical Physics |
| Leynes et al., 2018(Leynes et al. 2018) | Direct Generation of Pseudo-CT Images for Pelvic PET/MRI Attenuation Correction Using Deep Convolutional Neural Networks with Multiparametric MRI. | Journal of Nuclear Medicine |
| Xiang et al., 2018(Xiang et al. 2018) | Deep embedding convolutional neural network for synthesizing CT image from T1-Weighted MR image. | Medical Image Analysis |
| Dinkla et al., 2018(Dinkla et al. 2018) | MR-Only Brain Radiation Therapy: Dosimetric Evaluation of Synthetic CTs Generated by a Dilated Convolutional Neural Network. | International Journal of Radiation Oncology, Biology, Physics |
| Dinkla et al., 2019(Dinkla et al. 2019) | Dosimetric evaluation of synthetic CT for head and neck radiotherapy generated by a patch-based three-dimensional convolutional neural network. | Medical Physics |
| Wang et al., 2019(Wang et al. 2019) | Synthetic CT Generation Based on T2 Weighted MRI of Nasopharyngeal Carcinoma (NPC) Using a Deep Convolutional Neural Network (DCNN). | Frontiers in Oncology |
| Torrado-Carvajal et al., 2019(Torrado-Carvajal et al. 2019) | Dixon-VIBE Deep Learning (DIVIDE) Pseudo-CT Synthesis for Pelvis PET/MR Attenuation Correction. | Journal of Nuclear Medicine |
| Florkow et al., 2020(Florkow et al. 2020) | Deep learning-based MR-to-CT synthesis: The influence of varying gradient echo-based MR images as input channels. | Magnetic Resonance in Medicine |
| Gupta et al., 2019(Gupta et al. 2019) | Generation of Synthetic CT Images From MRI for Treatment Planning and Patient Positioning Using a 3-Channel U-Net Trained on Sagittal Images. | Frontiers in Oncology |
| Spadea et al., 2019(Spadea et al. 2019) | Deep Convolution Neural Network (DCNN) Multiplane Approach to Synthetic CT Generation From MR images-Application in Brain Proton Therapy. | International Journal of Radiation Oncology, Biology, Physics |
| Chen et al., 2018(Chen, Gong, et al. 2018) | Ultra-Low-Dose (18)F-Florbetaben Amyloid PET Imaging Using Deep Learning with Multi-Contrast MRI Inputs. | Radiology |

| Liu et al., 2019(Liu and Qi 2019) | Higher SNR PET image prediction using a deep learning model and MRI image. | Physics in Medicine & Biology |
|---|---|---|
| Liu et al., 2020(Liu et al. 2020) | Abdominal synthetic CT generation from MR Dixon images using a U-Net trained with 'semi-synthetic' CT data. | Physics in Medicine & Biology |
| Bahrami et al., 2020(Bahrami et al. 2020) | A new deep convolutional neural network design with efficient learning capability: Application to CT image synthesis from MRI. | Medical Physics |
| Song et al., 2020(Song et al. 2020a) | Super-Resolution PET Imaging Using Convolutional Neural Networks. | IEEE Transactions on Computational Imaging |
| Victor et al., 2021(Victor et al. 2021) | Magnetic resonance imaging–based synthetic computed tomography of the lumbar spine for surgical planning: a clinical proof-of-concept. | Neurosurgical Focus |
| **Generative Adversarial Networks with Convolutional Neural Networks** | | |
| Nie et al., 2018(Nie et al. 2018) | Medical Image Synthesis with Deep Convolutional Adversarial Networks. | IEEE Transactions on Biomedical Engineering |
| Emami et al., 2018(Emami et al. 2018) | Generating synthetic CTs from magnetic resonance images using generative adversarial networks. | Medical Physics |
| Maspero et al., 2018(Maspero et al. 2018) | Dose evaluation of fast synthetic-CT generation using a generative adversarial network for general pelvis MR-only radiotherapy. | Physics in Medicine & Biology |
| Peng et al., 2020(Peng et al. 2020) | Magnetic resonance-based synthetic computed tomography images generated using generative adversarial networks for nasopharyngeal carcinoma radiotherapy treatment planning. | Radiotherapy and Oncology |
| Yang et al., 2020(Yang, Qian, and Fan 2020) | An Indirect Multimodal Image Registration and Completion Method Guided by Image Synthesis. | Computer and Mathematical Methods in Medicine |
| Brou Boni et al., 2020(Brou Boni et al. 2020) | MR to CT synthesis with multicenter data in the pelvic area using a conditional generative adversarial network. | Physics in Medicine & Biology |
| Tie et al., 2020(Tie et al. 2020) | Pseudo-CT generation from multi-parametric MRI using a novel multi-channel multi-path conditional generative adversarial network for nasopharyngeal carcinoma patients. | Medical Physics |
| Fetty et al., 2020(Fetty et al. 2020) | Investigating conditional GAN performance with different generator architectures, an ensemble model, and different MR scanners for MR-sCT conversion. | Physics in Medicine & Biology |
| Kazemifar et al., 2020(Kazemifar et al. 2020) | Dosimetric evaluation of synthetic CT generated with GANs for MRI-only proton therapy treatment planning of brain tumors. | Journal of Applied Clinical Medical Physics |
| Tang et al., 2021(Tang et al. 2021) | Dosimetric evaluation of synthetic CT image generated using a neural network for MR-only brain radiotherapy. | Journal of Applied Clinical Medical Physics |
| Liu et al., 2021(Liu et al. 2021) | Synthetic dual-energy CT for MRI-only based proton therapy treatment planning using label-GAN. | Physics in Medicine & Biology |
| Jin et al., 2019(Jin et al. 2019) | Deep CT to MR Synthesis Using Paired and Unpaired Data. | Sensors (Basel) |
| Jiang et al., 2019(Jiang et al. 2019) | Cross-modality (CT-MRI) prior augmented deep learning for robust lung tumor segmentation from small MR datasets. | Medical Physics |

| Rubin et al., 2019(Rubin and Abulnaga 2019) | CT-To-MR Conditional Generative Adversarial Networks for Ischemic Stroke Lesion Segmentation. | MIDL 2019 Conference Paper |
|---|---|---|
| Chartsias et al., 2017(Chartsias et al. 2017) | Adversarial image synthesis for unpaired multi-modal cardiac data. | Paper presented at the International workshop on simulation and synthesis in medical imaging. |
| Zhang et al., 2018(Zhang, Yang, and Zheng 2018) | Translating and Segmenting Multimodal Medical Volumes with Cycle- and Shape-Consistency Generative Adversarial Network. | Paper presented at: Proceedings of the IEEE conference on computer vision and pattern Recognition2018. |
| Huo et al., 2019(Huo et al. 2019) | SynSeg-Net: Synthetic Segmentation Without Target Modality Ground Truth. | IEEE Transactions on Medical Imaging |
| Hiasa et al., 2018(Hiasa et al. 2018) | Cross-Modality Image Synthesis from Unpaired Data Using CycleGAN. | Paper presented at the Simulation and Synthesis in Medical Imaging, Cham |
| Wolterink et al., 2017(Wolterink et al. 2017) | Deep MR to CT Synthesis Using Unpaired Data. | Paper presented at the Simulation and Synthesis in Medical Imaging, Cham. |
| Huo et al., 2018(Huo et al. 2018) | Adversarial synthesis learning enables segmentation without target modality ground truth. | Paper presented at the 2018 IEEE 15th International Symposium on Biomedical Imaging |
| Yang et al., 2018(Yang et al. 2018b) | Deep Learning in Medical Image Analysis and Multimodal Learning for Clinical Decision Support. | Cham: Springer International Publishing. |
| Wang et al., 2018(Wang, Zhou, et al. 2018) | Locality Adaptive Multi-modality GANs for High-Quality PET Image Synthesis | Medical image computing and computer-assisted intervention |
| Lei et al., 2019(Lei et al. 2019) | MRI-only based synthetic CT generation using dense cycle consistent generative adversarial networks. | Medical Physics |
| Wei et al 2018 & 2019(Wei et al. 2018, 2019) | Learning Myelin Content in Multiple Sclerosis from Multimodal MRI Through Adversarial Training. | Paper presented at the Medical Image Computing and Computer Assisted Intervention – MICCAI 2018, Cham. |
| | Predicting PET-derived demyelination from multimodal MRI using sketcher-refiner adversarial training for multiple sclerosis. | Medical Image Analysis |
| Pan et al., 2018(Pan et al. 2018) | Synthesizing Missing PET from MRI with Cycle-consistent Generative Adversarial Networks for Alzheimer's Disease Diagnosis. | Paper presented at the Medical Image Computing and Computer Assisted Intervention – MICCAI 2018, Cham. |
| Choi et al., 2018(Choi and Lee 2018) | Generation of Structural MR Images from Amyloid PET: Application to MR-Less Quantification. | Journal of Nuclear Medicine |
| Bi et al., 2017(Bi et al. 2017) | Synthesis of Positron Emission Tomography (PET) Images via Multi-channel Generative Adversarial Networks (GANs). | Paper presented at the Molecular Imaging, Reconstruction and Analysis of Moving Body Organs, and Stroke |
| | | Imaging and Treatment, Cham. |
| Ben-Cohen et al., 2019(Ben-Cohen et al. 2019) | Cross-modality synthesis from CT to PET using FCN and GAN networks for improved automated lesion detection. | Engineering Applications of Artificial Intelligence, |
| Armanious et al., 2020(Armanious, Jiang, et al. 2020) | MedGAN: Medical image translation using GANs. | Computerized Medical Imaging and Graphics |
| Armanious et al., 2020(Armanious, Hepp, et al. 2020) | Independent attenuation correction of whole body [(18)F] FDG-PET using a deep learning approach with Generative Adversarial Networks. | EJNMMI Research |
| Vitale et al., 2020(Vitale et al. 2020) | Improving realism in patient-specific abdominal ultrasound simulation using CycleGANs | . International Journal of Computer Assisted Radiology and Surgery |
| Song et al., 2020(Song et al. 2020b) | PET image super-resolution using generative adversarial networks | Neural Networks |

## APPENDIX B: The CNN for cross-modality image estimation

A CNN is an artificial neural network (ANN) consisting of an input layer, one or more hidden layers, and an output layer. By definition, a CNN has a minimum of one convolutional layer (hidden) incorporated into the network. This layer performs a convolution operation on the input. The output of the convolution layer is a feature map generally of a smaller size and becomes the input to the next layer. Deep learning is an extension of the CNN framework catering for large datasets generally containing highly complex information in the input data, necessitating the need to develop large neural networks (Mamoshina et al. 2016).

CNNs typically start with a convolutional layer followed by activation functions. The subsequent layer might be a convolutional layer or a pooling layer for down-sampling the feature map. Depending on the CNN architecture, up-sampling layers, such as unpooling and transposed convolutional layers, may additionally be introduced into the network to enlarge feature map size. At the end of the network at least one fully connected layer is implemented to weight and determine the features, forming the CNN output. Within a CNN encoder-decoder framework (see U-Net shown in Figure 9), the encoder part has the action of dimensionality reduction into features, and the decoder up-samples features to an image.

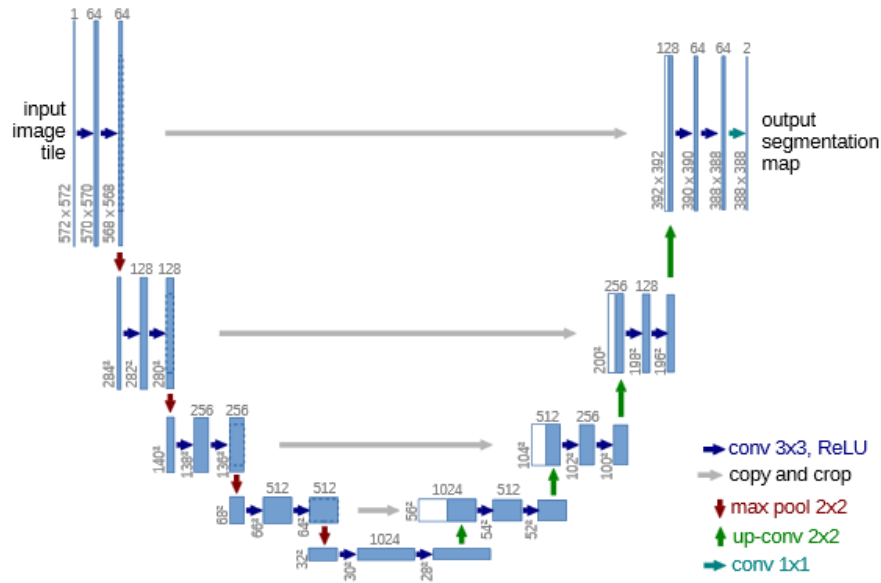

*Figure 9 Depict is the U-Net architecture. Blue boxes are multi-channel feature maps where information about the number of channels is provided on top of each box and the size of the feature maps are on their sides. The first half of the network is responsible for contracting (encoder) the feature maps through convolutional and pooling operations. The second half is responsible for expanding (decoder) the feature maps and resize them to the original size by concatenating the corresponding feature maps followed by convolutional operations. The U-Net autoencoder diagram has been adapted from(Ronneberger, Fischer, and Brox 2015).*

## B.1 Down-sampling layers

Convolutional layers mimic the convolution mathematical operation by applying a kernel filter to the input. Essentially, the convolution layer connection weights define the kernel, see U-Net and SegNet encoders in Figure 9 and Figure 10. The output of the layer is a feature map as defined by the kernel operation. In imaging applications, the filters are typically 2D or 3D matrices arranged based on the network dimensions.

Pooling layers perform a non-trainable operation and they are used to reduce the complexity in the input data by 'pooling' values (see U-Net and SegNet encoder in Figure 9 and Figure 10). The most common pooling operation is max pool, which returns the maximum value in a select pooling window. The reduction in data size is determined by the size of the pooling window.

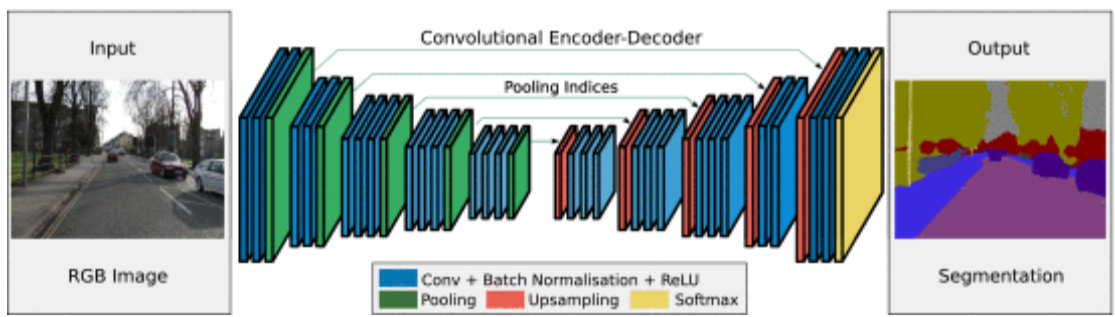

*Figure 10 SegNet architecture consists of encoder and decoder networks, each with 13 convolutional layers. Up-sampling is performed by pooling indices which are already computed during encoding and therefore it is transferred from the corresponding feature maps in encoder network. This encoder-decoder network has been adapted from (Badrinarayanan, Kendall, and Cipolla 2017).*

## B.2 Up-sampling layers

Within the CNN architecture, up-sampling layers are used similarly to down-sampling layers, but they instead increase output dimensionality. The unpooling layer (see decoder for SegNet in Figure 10), which is a non-trainable operation, projects a single value into a window via zero padding. The transposed convolutional layer will distribute a value into a window as defined by the kernel weights (see decoder for U-Net in Figure 9).

### B.3 Fully connected layers

The fully connected network, or FCN, uses fully connected layers at any point within a network. This layer provides a mechanism for making distal relationships in relation to the structure of the input. In imaging, for example, such a layer allows connections to be made between features in images at a distance away from each other. Fully connected layers also provide a way for retaining spatial relationships in the data when 2D or 3D images are converted to 1D vectors of image intensities and used as input to the CNN.

### B.4 Activation functions

Activation functions are predefined mathematical operations used to impose linear or nonlinear operations within the neural network. Output of the convolution operation is processed through an activation function before passed to the next layer. Different activation functions may be used after each layer. The choice of activation function in the network is very important and network performance changes with changes in activation function. A popular activation function in imaging applications is the rectified linear unit (ReLU), which maps a negative value to 0 (i.e. 0 for x < 0) and the positive value remains unchanged (i.e. x for x ≥ 0).

### B.5 CNN optimisation

The defined CNN architecture consists of layers, connections between layers and weights within layers. The number, type and size of layers is defined by the user. The weights within layers have to be determined through a process of CNN training and testing. This requires input data and matched output data for training the CNN. As such, application specific data for training is required. Ideally, the CNN prediction should match the expected output, and the difference is measured using a loss function. Different types of loss functions could be used (1-norm, 2-norm, etc.). Ideally, the value produced by the loss function reduces with epochs, i.e. training cycles. Once trained, that is when network weights have converged to specific values as determined by a sufficiently small loss function value, the CNN is evaluated using a testing dataset which is different to the training dataset. If the loss function performance is comparable between training and testing, then the CNN architecture and weights are said to be optimised. Otherwise, when training loss and testing losses follow different trends, either underfitting (not enough CNN architecture complexity to capture features in the data) or overfitting (too much CNN architecture complexity) may have occurred.

### APPENDIX C: The GAN for cross-modality image estimation

GAN is an extension of the CNN framework where two networks are trained simultaneously, a so-called generator and discriminator. The generator network is akin to those used in CNNs, for example the U-Net. Essentially, the first network performs image estimation and the discriminator takes the output of the CNN and checks whether it is a real image, or not. The two networks therefore compete against each other, and training is achieved when the discriminator loss is minimised. In other words, the discriminator cannot distinguish whether the image produced by the generator is fake.

A challenge of GANs is to achieve a balance between the generator and the discriminator. One may dominate the other, and it is likely to be the generator based on previous experience.(Yi, Walia, and Babyn 2019) Here, the discriminator cannot distinguish fake images produced by the generator from real images. Conversely, if the discriminator dominates, then the CNN generated samples are easily distinguishable from real images, and the discriminator training rate stalls, and consequently training of the generator ends. This is a common problem associated with estimating high resolution images, as high spatial variability in images challenge the network (Yi, Walia, and Babyn 2019). An additional consideration is mode collapse, defined as the generator producing limited sample variety, which can be the case when the network is insufficiently large for the input data.

## C.1 GAN architectures for cross-modality image estimation

In principle, the GAN frameworks provided in Table 6 can be applied to generate any modality imaging data. It is possible to change the GAN structure by introducing additional conditions for generating the output, referred to as conditional GAN (cGAN). Studies have shown the incorporation of conditions within the GAN framework leads to better image estimations (Dar et al. 2019). The cGAN layouts shown in Table 6 have been investigated for cross-modality image estimation:

i) VAEGAN - A variational autoencoder network (VAE) paired with a GAN in a way which allows sharing of parameters across the two networks within a joint training framework. For specifics on the method the reader is referred to two articles (Kingma and Welling 2013; Larsen et al. 2016).

ii) Pix2pix (for paired data) – A CNN, often the autoencoder type, is used as the generator and the discriminator checks how the estimated image resembles the actual image. The method was initially proposed by Isola et al. (Isola et al. 2017; Yi, Walia, and Babyn 2019). Here, a loss function specific to the pix2pix architecture is defined and assumed to be generalisable across different applications.

iii) CycleGAN (for unpaired data) – This type of GAN allows network training when there is no direct spatial correspondence between the input and target images, thus the standard adversarial loss is not able to differentiate between the two images. Originally, CycleGANs were applied in image-to-image translation applications, e.g. Zhu et al. (Zhu et al. 2017), effectively eliminating the need to align images prior to training. The cycle consistency loss was introduced to be able to train such a network. According to the study by Wolterink et al. on synthesising CT images from MRI (Wolterink et al. 2017), where both paired and unpaired datasets were available, the neural network trained with unpaired data resulted in a better MAE and PSNR.

iv) UNIT - An Unsupervised Image-to-image Translation (UNIT) neural network consisting of a coupled pair of VAEGANs (Liu, Breuel, and Kautz 2017). Each VAEGAN in a generator network accounts for a single imaging modality, and the neural network is constructed in a way that the latent space is shared. For more information about UNIT and the comparison between UNIT and CycleGAN we refer the readers to other publications (Jiang et al. 2019; Welander, Karlsson, and Eklund 2018).

Table 6. Shown are the conventional GAN and different types of conditional GAN used in cross-modality clinical image estimations.

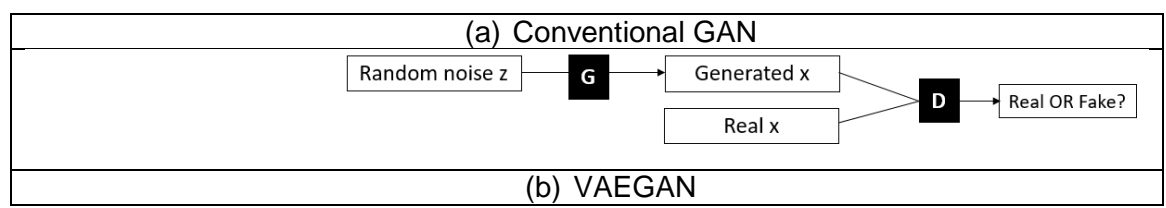

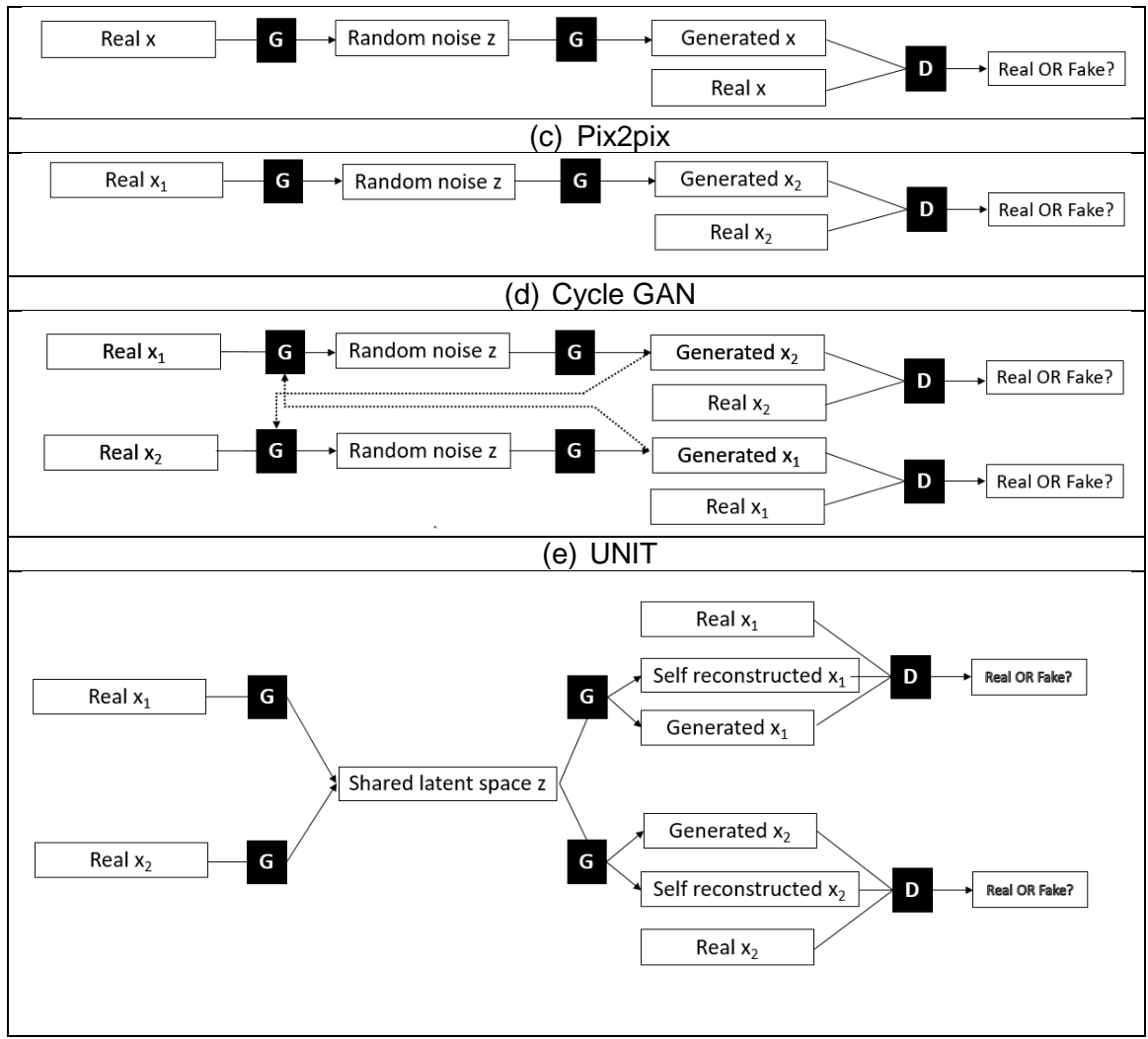

## 7. Vitae

Azin Shokraei Fard is a Master of Philosophy student at the Centre for advanced imaging, University of Queensland. She holds a bachelor's degree in Biomedical Engineering and Master's of Magnetic Resonance Technology. Her research focuses on artificial intelligence methods in molecular imaging and she is applying her image processing and machine learning expertise to implement cross-modality image estimation targeting neurological diseases and disorders.

Professor David Reutens is the inaugural director of the Centre for Advanced Imaging, University of Queensland. He is also a clinical neurologist specialising in epilepsy and is a senior staff specialist at the Royal Brisbane and Women's Hospital. Research in Professor Reuten's group focuses on neurological disorders, such as epilepsy, stroke and dementia, and the development of imaging methods to better understand, diagnose and manage them.

Associate Professor Viktor Vegh is the image analysis and methodology development group leader at the Centre for Advanced Imaging, University of Queensland. His interests lie in the research and development of medical imaging methods for the direct in vivo mapping of biological effects, translatable to the diagnosis and monitoring of neurological diseases and disorders. As such, his research focuses on understanding the underlying biological and physical processes influencing signal formation, specifically in magnetic resonance imaging.

## 8. Acknowledgements

Authors acknowledge funding from the Australian Research Council (IC170100035) in relation to the ARC Training Centre for Innovation in Biomedical Imaging Technology.